\documentclass[10pt]{article}
\usepackage{amsmath}
\usepackage{amssymb}
\usepackage{graphicx}
\usepackage{cite}
\usepackage{color}
\usepackage[labelfont=bf,labelsep=period,justification=raggedright]{caption}

\makeatletter
\renewcommand{\@biblabel}[1]{\quad#1.}
\makeatother

\date{}

\begin{document}

\begin{flushleft}
{\Large
\textbf{Navigating the massive world of reddit: Using backbone networks to map user interests in social media}
}
\\
Randal S.~Olson$^{1,\ast}$, 
Zachary P. Neal$^{2}$
\\
\bf{1} Department of Computer Science \& Engineering
\\
\bf{2} Department of Sociology
\\
Michigan State University, East Lansing, MI 48824, U.S.A.
\\
$\ast$ E-mail: olsonran@msu.edu
\end{flushleft}

\section*{Abstract}
In the massive online worlds of social media, users frequently rely on organizing themselves around specific topics of interest to find and engage with like-minded people. However, navigating these massive worlds and finding topics of specific interest often proves difficult because the worlds are mostly organized haphazardly, leaving users to find relevant interests by word of mouth or using a basic search feature. Here, we report on a method using the backbone of a network to create a map of the primary topics of interest in any social network. To demonstrate the method, we build an interest map for the social news web site reddit and show how such a map could be used to navigate a social media world. Moreover, we analyze the network properties of the reddit social network and find that it has a scale-free, small-world, and modular community structure, much like other online social networks such as Facebook and Twitter. We suggest that the integration of interest maps into popular social media platforms will assist users in organizing themselves into more specific interest groups, which will help alleviate the overcrowding effect often observed in large online communities.

\section*{Introduction}
In the past decade, social media platforms have grown from a pastime for teenagers into tools that pervade nearly all modern adults' lives~\cite{Rainie2012}. Social media users typically organize themselves around specific interests, such as a sports team or hobby, which facilitates interactions with other users who share similar interests. For example, Facebook users subscribe to topic-specific ``pages''~\cite{Strand2011}, Twitter users classify their tweets using topic-specific ``hashtags''~\cite{Chang2010}, and reddit users post and subscribe to topic-specific sub-forums called ``subreddits''~\cite{Gilbert2013}. 

These interest-based devices provide structure to the growing worlds of social media, and are essential for the long-term success of social media platforms because they make these big worlds feel small and navigable. However, navigation of social media is challenging because these worlds do not come with maps~\cite{Boguna2008, Benevenuto2012}. Users are often left to discover pages, hashtags, or subreddits of interest haphazardly, by word of mouth, following other users' ``votes'' or ``likes'', or by using a basic search feature. Owing to the scale-free structure of most online social networks, these elementary navigation strategies result in users being funnelled into a few large and broad interest groups, while failing to discover more specific groups that may be of greater interest~\cite{Albert1999, Barabasi2000}.

In this work, we combine techniques for network backbone extraction and community detection to construct a roadmap that can assist social media users in navigating these interest groups by identifying related interest groups and suggesting them to users. We implement this method for the social news web site reddit~\cite{reddit-what-is}, one of the most visited social media platforms on the web~\cite{reddit-alexa-ranking}, and produce an interactive map of all of the subreddits. An interactive version of the reddit interest map is available online~\cite{redditviz}.

By viewing subreddits as nodes linked by users with common interests, we find that the reddit social media world has a scale-free, small-world, and modular community structure. The scale-free property is the expected outcome of a preferential attachment process and helps explain the challenges of haphazard navigation. Additionally, the small-world property explains how the big world of reddit can seem small and navigable to users when it is mapped out. Finally, the modular community structure in which narrow interest-based subreddits (e.g., dubstep or rock music) are organized into broader communities (e.g., music) allows users to easily identify related interests by zooming in on a broader community. We suggest that the integration of such interest maps into popular social media platforms will assist users in organizing themselves into more specific interest groups, which will help alleviate the overcrowding effect often observed in large online communities~\cite{Gilbert2013}.

Further, this work releases and provides an overview of a data set of over 850,000 anonymized reddit user's interests, thus establishing another standard real-world social network data set for researchers to study. This is useful because, although reddit is among the largest online social networks and has been identified as a starting point for the viral spread of memes and other online information~\cite{Sanderson2013}, it has been relatively understudied~\cite{Wasike2011, Merritt2012, Gilbert2013}. This data set can be downloaded online at~\cite{reddit-data}.

\begin{figure}[ht]
\begin{center}
\includegraphics[width=0.9\textwidth]{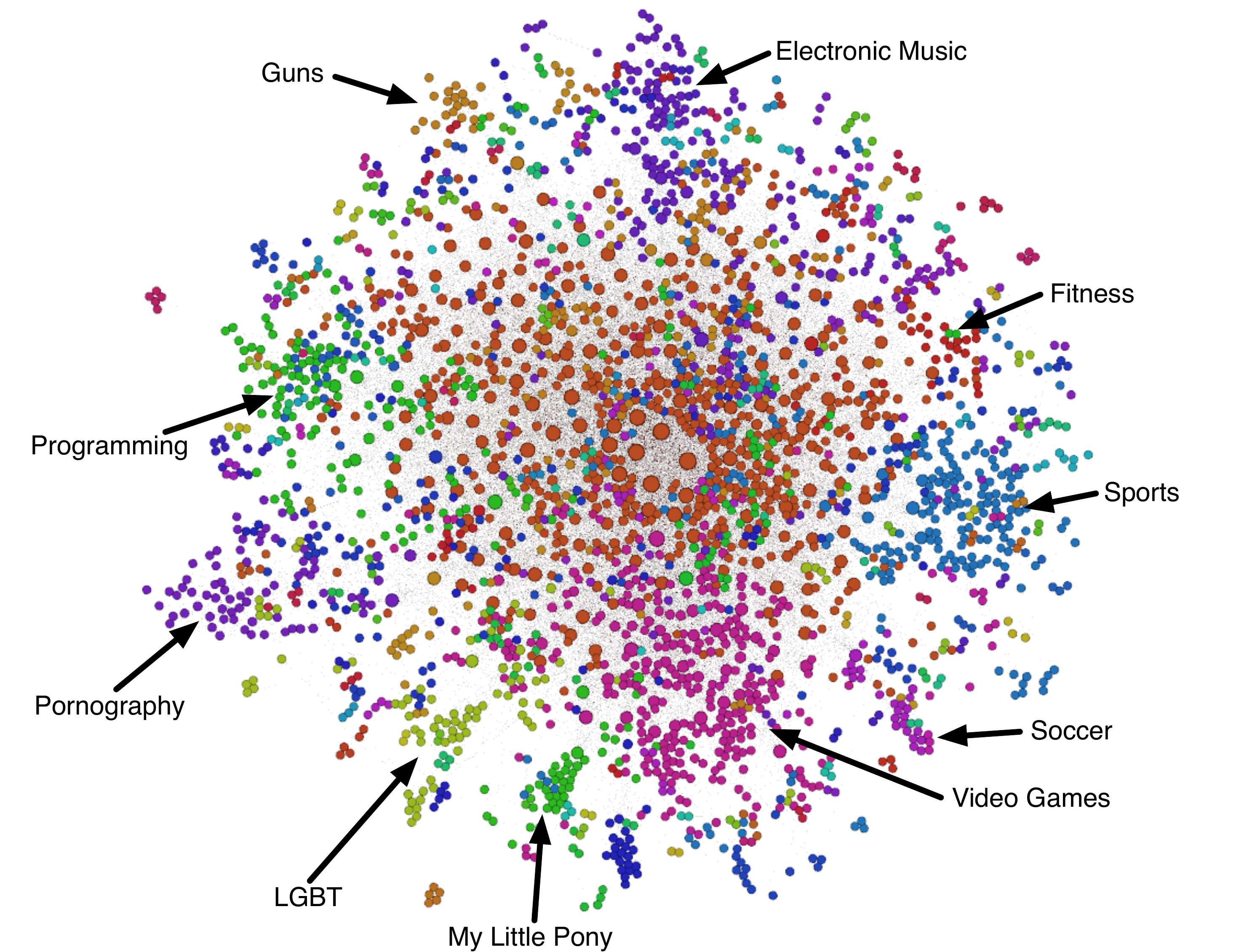}
\end{center}
\caption{
{\bf Reddit interest network.} The largest components of the reddit interest network is shown with 10 interest meta-communities annotated; it closely matches the structure of other online social networks including Flikr and Yahoo360~\cite{Kumar2010}. Each node is a single subreddit, where color indicates the interest meta-community that the subreddit is a member of. Nodes are sized by their weighted PageRank to provide an indication of how likely a node is to be visited, and positioned according to the OpenOrd layout in Gephi to place related nodes together. An interactive version of the reddit interest map is available online at \url{http://rhiever.github.io/redditviz/clustered/}}
\label{fig:full-interest-network}
\end{figure}

\section*{Results}
\subsection*{Reddit interest map}
For the final version of the reddit interest map, we use the backbone network produced with $\alpha = 0.05$ (see Methods). This results in a network with 59 distinct clusters, which we call {\it interest meta-communities}. In Figure~\ref{fig:full-interest-network}, the nodes (i.e., subreddits) are sized by their weighted PageRank~\cite{Page1999} to provide an indication of how likely a node is to be visited, and positioned according to the OpenOrd layout in Gephi~\cite{Bastian2009} to place related nodes together.

\begin{figure}
\begin{center}
\includegraphics[width=0.8\textwidth]{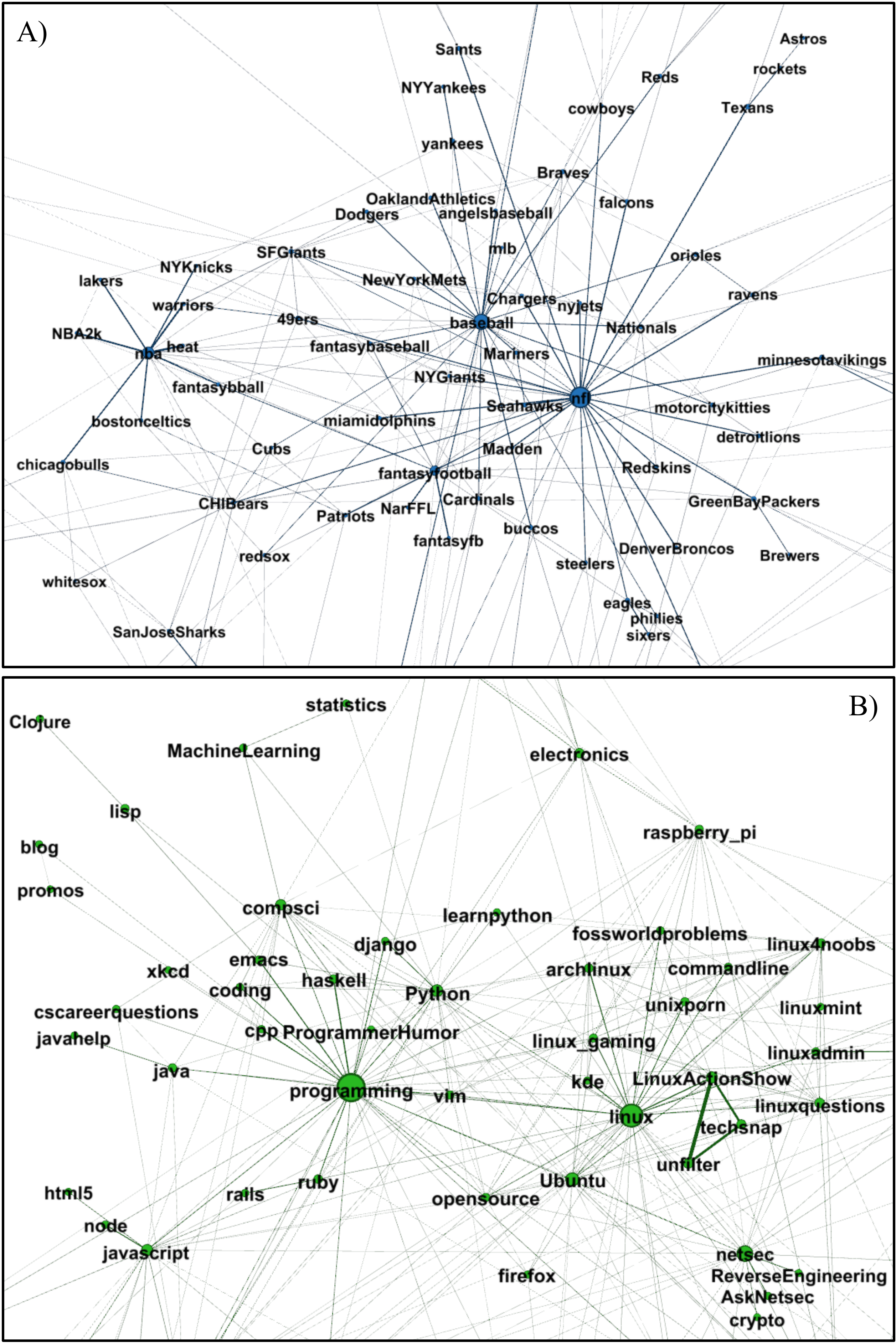}
\end{center}
\caption{
{\bf Example reddit interest meta-communities.} Pictured are several topic-specific subreddits composing a meta-community around a broad topic such as sports (A) or programming (B). Each node is a subreddit, and each edge indicates that a significant portion of the posters in the two subreddits post in both subreddits (see Methods).}
\label{fig:interest-communities}
\end{figure}

Through this method, we immediately see several distinct interest meta-communities, 10 of which are annotated in Figure~\ref{fig:full-interest-network}. These interest meta-communities act as starting points in the interest map to show the broad interest categories that the entire reddit community is discussing. From these starting points, users can zoom in on a single broad interest category to find subreddits dedicated to more specific interests, as shown in Figure~\ref{fig:interest-communities}. Notably, there is a large, orange interest meta-community in the center of the interest map that overlaps with several other interest meta-communities. This orange interest meta-community represents the most popular, general interest subreddits (e.g., ``pictures'' and ``videos'') in which users of all backgrounds regularly participate, and thus are expected to have considerable overlap with many other communities.

Figure~\ref{fig:interest-communities} depicts zoomed-in views of two interest meta-communities annotated in Figure~\ref{fig:full-interest-network}. In Figure~\ref{fig:interest-communities}A, the ``sports'' meta-community, specific sports teams are organized around the corresponding sport that the teams play in. For example, subreddits dedicated to discussion of the Washington Redskins or Denver Broncos -- relatively small, specific subreddits -- are organized around the larger, more general interest NFL subreddit where users discuss the latest NFL news and games. Similarly in Figure~\ref{fig:interest-communities}B, the ``programming'' meta-community, subreddits dedicated to discussing programming languages such as Python and Java are organized around a more general programming subreddit, where users discuss more general programming topics.

This backbone network structure naturally lends itself to an intuitive interest recommendation system. Instead of requiring a user to provide prior information about their interests, the interest map provides a hierarchical view of all user interests in the social network. Further, instead of only suggesting interests immediately related to the user's current interest(s), the interest map recommends interests that are potentially two or more links away. For example in Figure~\ref{fig:interest-communities}A, although the Miami Heat and Miami Dolphins subreddits are not linked, Miami Heat fans may also be fans of the Miami Dolphins. A traditional recommendation system would only recommend NBA to a Miami Heat fan, whereas the interest map also recommends the Miami Dolphins subreddit because they are members of the same interest meta-community.

\subsection*{Network properties}

\begin{figure}
\begin{center}
\includegraphics[width=\textwidth]{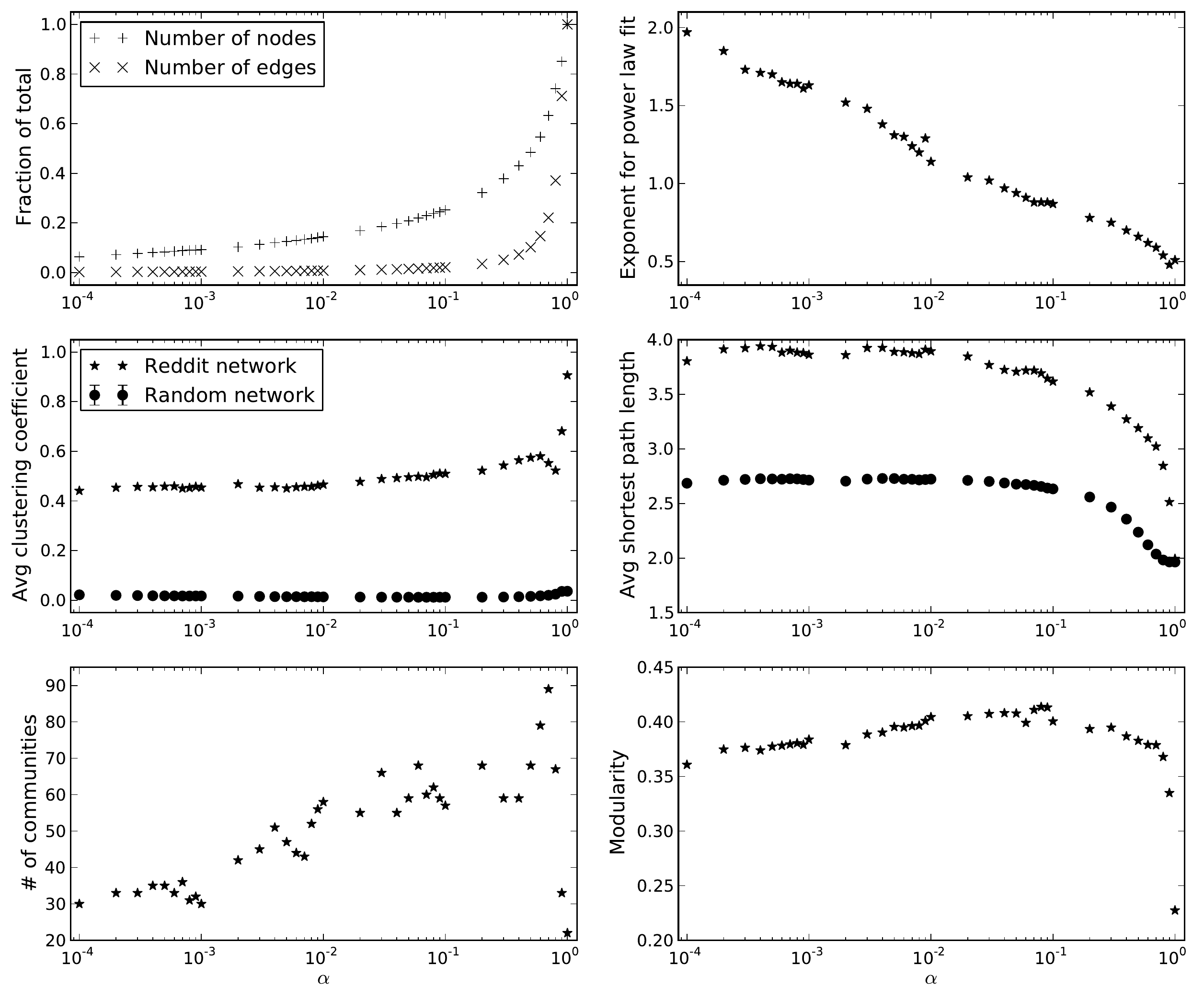}
\end{center}
\caption{
{\bf Network statistics for the backbone network.} Sensitivity analysis of the reddit interest network over a range of $\alpha$ cutoff values. Lower $\alpha$ means that fewer statistically significant edges are pruned. In general, this sensitivity analysis shows that the backbone interest network is stable for $\alpha$ cutoff values $\leq 0.05$. Error bars for the Erd\H{o}s-–R\'{e}nyi random networks are two standard deviations over 30 random networks, and are too small to show up on the graph. Note the logarithmic scale of the x-axis.}
\label{fig:bothsig_stats}
\end{figure}

In Figure~\ref{fig:bothsig_stats}, we show a series of network statistics to provide an overview of the backbone reddit interest network. These network statistics are plotted over a range of $\alpha$ cutoff values for the backbone reddit interest network (see Methods) to demonstrate that the interest network we chose in Figure~\ref{fig:full-interest-network} is robust to relevant $\alpha$ cutoff values.

As expected, the majority of the edges are pruned by an $\alpha$ cutoff of 0.05 (Figure~\ref{fig:bothsig_stats}, top left). This result demonstrates that the backbone interest network is stable with an $\alpha$ cutoff $\leq 0.05$, which is the most relevant range of $\alpha$ cutoffs to explore. Surprisingly, 80\% of the subreddits that we investigated -- roughly 12,000 subreddits -- do not have enough users that consistently post in another subreddit to maintain even a single edge with another subreddit. The majority of these 12,000 subreddits likely do not have any significant edges due to user inactivity, e.g., some subreddits have only a single user that frequently posts to them (Table~\ref{tab:network_desc}). Another factor that likely contributes to the 12,000 unlinked subreddits is temporary interests, i.e., an interest such as the U.S. Presidential election that temporarily draws a large number of people together, but eventually fades into obscurity again.

Next, we are interested in exploring whether the backbone reddit interest network is a scale-free network, where preferential attachment to subreddits results in a few extremely popular (i.e., connected) subreddits and mostly unpopular subreddits. As such, scale-free networks are known to have node degree distributions that fit a power law~\cite{Albert1999, Barabasi2000}. Regardless of the $\alpha$ cutoff, we observed that the node degree distribution of all backbone reddit interest networks fit a power law (${\rm R^2\approx0.91}$ for $k\geq50$; Figure~\ref{fig:bothsig_stats}, top right). This scale-free network structure is likely partially due to reddit's default subreddit system~\cite{reddit-defaults}, where newly registered users are subscribed to a set of 20 subreddits by default.

Furthermore, we want to confirm that the backbone reddit interest network is a small-world network~\cite{Barabasi1999}. Small-world networks are known to contain numerous clusters, as indicated by a high average clustering coefficient, with sparse edges between those clusters, which results in an average shortest path length between all nodes (${\rm L_{sw}}$) that scales logarithmically with the number of nodes (N):

\begin{equation}
{\rm L_{sw} \approx log_{10}(N)}
\label{eq:length-sw}
\end{equation}

Figure~\ref{fig:bothsig_stats} (middle left and middle right) depicts the average clustering coefficient and shortest path length for all nodes in the backbone reddit interest network. Compared to Erd\H{o}s-–R\'{e}nyi random networks with the same number of nodes and edges, the backbone network has a significantly higher average clustering coefficient. Similarly, the measured average shortest path length of the backbone network ($\alpha$ cutoff $= 0.05$) follows Equation~\ref{eq:length-sw}, with ${\rm L_{sw} = log_{10}(2,347) = 3.37} \approx 3.71$ from Figure~\ref{fig:bothsig_stats} (middle right). Thus, the backbone reddit interest backbone network qualitatively appears to exhibit small-world network properties.

To quantitatively determine whether the reddit interest network exhibits small-world network properties, we used the small-worldness score (${\rm S_G}$) proposed in~\cite{Humphries2008}:

\begin{equation}
{\rm S_G = \frac{{C_G}/{C_{rand}}}{{L_G}/{L_{rand}}}}
\end{equation}
where C is the average clustering coefficient, L is the average shortest path length between all nodes, G is the network the small-worldness score is being computed for, and ``rand'' is an Erd\H{o}s-–R\'{e}nyi random network with the same number of nodes and edges as G. If ${\rm S_G} > 1$, then the network is classified as a small-world network. For the backbone reddit network, we calculated ${\rm S_G} = 14.2$ (${\rm P} < 0.001$), which indicates that the reddit interest network exhibits small-world network properties.


Now that we know that the backbone reddit interest network is scale-free and exhibits small-world network properties, we want to study the community structure of the backbone network. Shown in Figure~\ref{fig:bothsig_stats} (bottom right), the backbone network exhibits a consistently high modularity score with an $\alpha$ cutoff as high as 0.9, implying that even a slight reduction in the number of edges in the backbone network reveals the reddit interest community structure. Correspondingly, depicted in Figure~\ref{fig:bothsig_stats} (bottom left), the number of identified communities (i.e., clusters) remains relatively low until the $\alpha$ cutoff is reduced to $\leq 0.9$. As the $\alpha$ cutoff is reduced, the number of identified communities generally decreases, which coincides with the loss of nodes as $\alpha$ decreases. Thus, the backbone reddit interest network has $\approx 30$ core communities, and another $\approx 30$ weakly linked communities that are lost as a more stringent $\alpha$ cutoff is applied.

\section*{Discussion}
We have shown that backbone networks can be used to map and navigate massive interest networks in social media. By viewing the big world of reddit as a hierarchical map, users can now explore related interests without providing any prior information about their own interests. Future applications of this method may also facilitate navigation of other popular social network platforms such as Facebook and Twitter.

Furthermore, such an interest map could allow social media users to self-organize into more specific interest forums, thus reducing preferential attachment to large, general interest forums and alleviating the issues that arise in overcrowded social network forums~\cite{Gilbert2013}. Given previous work that suggests network properties such as small-worldness and even modularity can result solely from network growth processes~\cite{Hintze2010}, it would be interesting in future work to observe what processes govern network growth when users have access to an interests map like those shown in Figures~\ref{fig:full-interest-network} and ~\ref{fig:interest-communities}, and what network properties emerge from these growth processes.

This work provides a unique view of reddit that debunks a common misconception of the social news web site. Typically, outsiders view reddit as a single, homogeneous entity that acts as one, e.g. ``Should Reddit Be Blamed for the Spreading of a Smear?''~\cite{reddit-news}. In contrast, the reddit interest map shown here provides a different view of reddit, where many users organize themselves into cliques based on shared interests and rarely interact with other reddit users outside their clique. In that light, we hope this work reveals that, like many social communities (online or offline), reddit is a community composed of a diverse group of people that are brought together by thousands of seemingly-unrelated interests.

Additionally, we explored the network properties of the backbone reddit interest network that we composed from the posting behavior of over 850,000 active reddit users. In this analysis, we found that the reddit interest network has a scale-free, small-world, and modular community structure, corroborating findings in many other online social networks~\cite{Ahn2007, Mislove2007}. Uniquely, reddit potentially enforces a scale-free network structure on its users by automatically subscribing all new users to the same set of 20 subreddits~\cite{reddit-defaults}. Exploring the effect of automatically subscribing users to a fixed set of interest-specific forums on social interest network structure could be another interesting venue of future work. To expedite future analyses of the reddit interest network, we have provided the raw, anonymized data set available to download online~\cite{reddit-data}.

It is important to note that the sample of user behavior we have taken is cross-sectional, reflecting users' reddit posts and thus the relationships among reddit interests at a fixed point in time in mid-2013. However, as users' interests evolve, so too do the relationships among them~\cite{Banerjee2009}. In some cases, highly specialized and related subreddits may fuse into a single subreddit, while in other cases a general subreddit may split into multiple more specialized ones. Thus, such an interest map would require periodic (or, ideally, real-time) updating to accurately reflect dominant interests in the social network and their relationships to one another.

\section*{Methods}
To acquire the data for this study, we mined user posting behavior data from reddit by first gathering the user names of 876,961 active users that post to 15,122 distinct subreddits (see Table~\ref{tab:data_desc} for more detail). We note that reddit reports to have over 2.6 million registered users as of December 2013~\cite{reddit-about}, so this data set represents a random sample of roughly 1/3 of the total active users on reddit. For each of the users, we gathered their 1,000 most recent link submissions and comments, counted how many times they post to each subreddit, and registered them as interested in a subreddit only if they posted there at least 10 times. We applied this threshold of at least 10 posts to filter out users that are not active in a particular subreddit.

From these data, we defined a bipartite network $\mathbf{X}$, where $X_{ij} = 1$ if user $i$ is an active poster in subreddit $j$ and otherwise is 0. We then projected this as a weighted unipartite network $\mathbf{Y}$ as $\mathbf{XX'}$, where $Y_{ij}$ is the number of users that post in both subreddits $i$ and $j$. This resulted in 4,520,054 non-zero edges between the subreddits. Details of the raw weighted subreddit network are shown in Table~\ref{tab:network_desc}.

\begin{table}[h]
\centering
\caption{
\bf{Edge weights in the raw and backbone reddit interest networks}}
\begin{tabular*}{0.45\columnwidth}{|c|c c c|}
\hline
Network & Mean & Minimum & Maximum \\
\hline
Raw & 17.77 & 1 & 309,985 \\
Backbone & 0.0052 & 0.00068 & 0.1997\\
\hline
\end{tabular*}
\label{tab:network_desc}
\end{table}

Due to the challenges associated with analyzing large weighted networks, we reduced the number of edges in the weighted subreddit network using a backbone extraction algorithm~\cite{Serrano2009}. This backbone extraction algorithm preserves edges whose weight is statistically incompatible, at a given level of significance $\alpha$, with a null model in which edge weights are distributed uniformly at random. In the resulting backbone network, two subreddits are linked if the number of users who post in both of them is statistically significantly larger than expected in a null model, from the perspective of \emph{both} subreddits. To combine the directed edges between each two nodes, we replaced the two directed edges with a single undirected edge whose weight is the average of the two directed edges.

Thus, this technique defines a network of subreddit pathways along which there is a high probability users might traverse if they navigate reddit by following the posts of other users. Adjusting the $\alpha$ parameter allows the backbone network to include more (e.g., when $\alpha$ if larger) or fewer (e.g., when $\alpha$ is smaller) such pathways. Figure~\ref{fig:bothsig_stats} summarizes the topological properties of backbones extracted using a range of $\alpha$ parameter values; in the findings and discussion we focus on a backbone extracted using the conventional $\alpha = 0.05$.

We used Python's PRAW package\footnote{Python Reddit API Wrapper (PRAW): \url{https://github.com/praw-dev/praw}} to gather the data and Python's NetworkX package~\cite{Hagberg2008} to compute all network statistics. In the backbone graph, we focus only on the largest connected component. We detected network communities using~\cite{Blondel2008} and visualized the communities using the OpenOrd node layout, both as implemented in Gephi~\cite{Bastian2009}.

\section*{Acknowledgments}
We gratefully acknowledge the support of the Michigan State University High Performance Computing Center and the Institute for Cyber Enabled Research (iCER). We thank Arend Hintze, Christoph Adami, and Emily Weigel for helpful feedback during the preparation of this manuscript.


\bibliographystyle{plos2009}
\bibliography{references}

\newpage
\section*{Supplementary Information}

\makeatletter 
\renewcommand{\thetable}{S\@arabic\c@table}
\makeatother
\setcounter{table}{0}

\begin{table}[h]
\centering
\caption{
\bf{Descriptive statistics of the bipartite (user-to-subreddit) network}}
\begin{tabular*}{0.465\columnwidth}{|c|c|}
\hline
Statistic & Value\\
\hline
Total \# of users & 876,961\\
Total \# of subreddits & 15,122\\
Average \# of subreddits per user & 9.69\\
Minimum \# of subreddits per user & 1\\
Maximum \# of subreddits per user & 112\\
Average \# of users per subreddit & 561.8\\
Minimum \# of users per subreddit & 1\\
Maximum \# of users per subreddit & 523,025\\
\hline
\end{tabular*}
\label{tab:data_desc}
\end{table}

\end{document}